\documentclass[conference]{IEEEtran}
\usepackage{amsmath}
\IEEEoverridecommandlockouts
\usepackage{cite}
\usepackage{amsmath,amssymb,amsfonts}
\usepackage{algorithmic}
\usepackage{graphicx}
\usepackage{textcomp}
\usepackage{xcolor}
\usepackage{subfig}

\def\minwrt[#1]{\underset{#1}{\mathrm{minimize }}}
\def\argminwrt[#1]{\underset{#1}{\text{arg min }}}

\def\BibTeX{{\rm B\kern-.05em{\sc i\kern-.025em b}\kern-.08em
    T\kern-.1667em\lower.7ex\hbox{E}\kern-.125emX}}
\begin{document}

\title{Room Impulse Response Estimation through Optimal Mass Transport Barycenters\\}

\author{
	\IEEEauthorblockN{Rumeshika Pallewela\textsuperscript{+}, Yuyang Liu\textsuperscript{+} and Filip Elvander \\
	}
	\IEEEauthorblockA{ Dept. of Information and Communications Engineering, Aalto University, Finland \\
	Email: firstname.lastname@aalto.fi}\thanks{This research was supported in part by the Research Council of Finland
(decision number 362787).}
\thanks{+ Authors 1 and 2 have made equal contributions to this paper.}

}

\maketitle

\begin{abstract}
In this work, we consider the problem of jointly estimating a set of room impulse responses (RIRs) corresponding to closely spaced microphones. The accurate estimation of RIRs is crucial in acoustic applications such as speech enhancement, noise cancellation, and auralization. However, real-world constraints such as short excitation signals, low signal-to-noise ratios, and poor spectral excitation, often render the estimation problem ill-posed. In this paper, we address these challenges by means of optimal mass transport (OMT) regularization. In particular, we propose to use an OMT barycenter, or generalized mean, as a mechanism for information sharing between the microphones. This allows us to quantify and exploit similarities in the delay-structures between the different microphones without having to impose rigid assumptions on the room acoustics. The resulting estimator is formulated in terms of the solution to a convex optimization problem which can be implemented using standard solvers. In numerical examples, we demonstrate the potential of the proposed method in addressing otherwise ill-conditioned estimation scenarios.


\end{abstract}


\section{Introduction}
%
%
When modeling the acoustics of a room as a linear time-invariant system, the room impulse response (RIR) describes the impact of the acoustic environment on a sound signal when emitted and received by a point source and a point receiver.
%
Having access to an accurate estimate of an RIR is essential for applications such as noise cancellation~\cite{763310}, speech enhancement~\cite{vincent2018audio}, and auralization~\cite{kleiner1993auralization-an}.
The conventional approach to obtaining the RIR involves extensive measurements using a compact device and a broadband excitation signal~\cite{koyama2021meshrir}. However, this method is both time-consuming and costly, especially as the RIR is only a point-to-point description of the acoustic environment. Alternatively, one could envision estimating an RIR using ambient signals such as, speech. Nevertheless, the accuracy of an RIR estimate, or even the feasibility of obtaining an estimate, depends on the duration and spectral characteristics of the excitation signal, as well as on the presence of noise. In order to address this,
%
%
various model-based estimation methods have been proposed, often incorporating regularization techniques aiming to capture assumed features of the RIR. These include methods building on Tikhonov regularization, which can be understood as a Bayesian estimation technique~\cite{calvetti2000tikhonov, van2008optimally}, as well as sparsity-based methods originating from the compressed sensing framework (see, e.g.,~\cite{antonello2017room,crocco2015room,mignot2013low, mignot2013room,verburg2018reconstruction}). There are also works on using low-rank structures for modeling and estimating RIRs, motivated by steady-state descriptions of room modes~\cite{lowrankRIRestimation,lowranktensorRIRmodel} and spatio-temporal correlation ~\cite{kroneckerproductdecomp}.

In this work, we explore a different type of prior assumption on the structure of RIRs, namely that RIRs corresponding to closely-spaced sensors should be \textit{similar}. This similarity, if it can be well-defined and quantified, can then be used to jointly estimate the set of RIRs corresponding to, e.g., a microphone array. Herein, we propose to define this similarity between RIRs based on the concept of optimal mass transport (OMT). The problem of OMT~\cite{villani2009optimal} is concerned with how to best morph, or transport, one distribution of mass as to become identical to another. The resulting minimal cost of transport can then serve as a measure of similarity between the two distributions. Recently, this idea has been used for interpolating and estimating RIRs by considering an RIR to be a (signed) distribution over delay-space~\cite{MovingsourceRIRestDavidICASSP, OMTRIRestDavidAntonEusipco} or a distribution over 3D space for the case of spatial RIRs~\cite{geldert2023interpolation}. In~\cite{OMTRIRestDavidAntonEusipco}, an OMT regularizer was used for incorporating an approximate knowledge of the room geometry when inferring the RIR, whereas in~\cite{MovingsourceRIRestDavidICASSP}, the work considered estimating the RIR of a source moving on a smooth trajectory. 

Herein, we build on the later approach and propose a method for jointly estimating a set of RIRs corresponding to a microphone array. Under the assumption that the delay path from the source to a microphone varies smoothly under perturbations of the microphone location, we propose to express this assumed similarity in terms of an OMT distance and to exploit this in estimation by constructing an OMT barycenter. The barycenter can be understood as a generalized mean in the corresponding OMT distance and serves as a mechanism for regularization and information sharing between the different RIRs. Numerical simulations demonstrate that the proposed OMT barycenter method generally outperforms alternative estimation techniques, showcasing its robustness even under challenging conditions. 

\section{Signal model}\label{sigmod}

Consider a scenario in which a known source signal $x$ impinges on a set of $K \in \mathbb{N}$ microphones, yielding the observed signals
\begin{align} \label{eq:measurement_convolution}
    y_k(n) = (x*h_k)(n) + w_k(n), 
\end{align}
for $n = 0,1,\ldots,N \in \mathbb{N}$, and $k = 1,2,\ldots,K$. Here, $h_k$ denotes the RIR for the $k$th microphone, $x*h_k$ denotes the convolution of $x$ with $h_k$, and $w_k$ is an additive noise. Herein, we will assume that the additive noise is temporally and spatially white. For a more succinct notation, we will throughout use the equivalent representation
\begin{align}\label{eq:measurement_matrix}
    \mathbf{y}_k = \mathbf{X} \mathbf{h}_k + \mathbf{w}_k, \; k = 1,\ldots,K,
\end{align}
where $\mathbf{y}_k$, $\mathbf{h}_k$, and $\mathbf{w}_k$ are the vector representations of $y_k$, $h_k$, and $w_k$, respectively, and $\mathbf{X}$ is the matrix representation of the convolution with $x$.

Since we know the source signal $x$ and can measure its outputs $\mathbf{y}_k$ at each microphone, we want to recover how the room transforms $x$ on its way to each sensor. That is given by the RIRs $\mathbf{h}_k$, where the system follows a linear model with $\mathbf{h}_k$ as the unknown to be estimated. 

 However, even if $\mathbf{X}$ is full rank, estimating a \emph{single} RIR via standard methods can still be poorly conditioned when the source signal lacks sufficient spectral excitation, especially in narrowband scenarios like speech, rendering the estimation highly sensitive to measurement noise. In such cases, using a \emph{set} of these RIRs (e.g., measured under different positions or signals) can broaden the effective frequency coverage and thus yielding more robust estimates of $\mathbf{h}_k$.  
 
Moreover, from a time domain perspective, a measured RIR often exhibits distinct peaks corresponding to the direct path and the reflections from wall bounds and obstacles in the room, which are spaced according to their respective time delays. Accurately capturing these peaks is essential for acoustic measurements and for physically motivated models of the room response. One common model to simulate these time-domain reflection paths is the image source model (ISM). It is used to represent the  geometrical acoustics (GA) principles~\cite{Allen1979} and RIR $\mathbf{h}$, which can be denoted as a delay-magnitude pair based on ISM as follows.
\begin{equation}
\mathbf{h} = \{( {\tau_{m}, p_{m}} ) \in \mathbb{R} \mathrm{x} \mathbb{R}\} ,
\end{equation}
where $\tau_{m}$ denotes the delay related to the $m$th  image source represented by $|\Delta\mathbf{ d|}/c$ with $|\Delta\mathbf{ d|}$ denoting the distance between the image source and receiver. The variable ${p_{m}}$ denotes the pressure contribution to the $m$th filter tap of vector $\mathbf{h}$ and $p_{m} \propto \frac{1}{|\Delta\mathbf{ d|}}$.

The early part of  $\mathbf{h}$ corresponds to the reflection from the walls and  it can be modeled as originating  from an equivalent image source position. In our work, the radius of the microphone array is assumed to be much smaller than the dimensions of the room, causing the sensors to be positioned closely together. Particularly, in this situation, the delay amplitude pairs of the filter taps in RIRs remain similar to each other, drifting only by small time shifts. Therefore, their horizontal properties remain the same and we can expect the positions of the taps of $\mathbf{h}_{i}$ to be similar to $\mathbf{h}_{i+1}$. Structured around this assumption, we are able to utilize the closely spaced sensors to get a good estimate for a set of RIRs. In order to exploit this idea, we propose to utilize an important mathematical concept, namely, OMT. 


\subsection{OMT and barycenter formulation}\label{sec:OMT}

OMT is a mathematical framework that compares two mass distributions~\cite{peyre2019computational,villani2009optimal,SensorFusionFilipIcassp}. More specifically, it determines the optimal (least costly) way to transport one distribution to another. This minimum cost defines a distance between the distributions, inducing a geometric structure on the delay space of mass distributions. Consequently, this geometric framework enables estimation~\cite{10095082}, interpolation~\cite{estimationOT}, and the computation of generalized averages known as barycenters within the delay space, providing a rigorous method for comparing and combining mass distributions.


In our work, we are utilizing this framework to map each measured RIR  $\mathbf{h}_k$ of all the closely placed $K$ number of sensors in the array to a virtual reference RIR $\mathbf{h}_0$. By defining a minimum value for each distance using discrete  Monge - Kantrovich OMT problem, we can formulate the problem as follows,

\begin{equation}\label{eq:dotcal}
\begin{aligned}
\operatorname{dOT}(\mathbf{h}_0, \mathbf{h}_k) &= \min_{\mathbf{M}_k\geq 0} \langle \mathbf{C}, \mathbf{M}_k \rangle \\
&\quad \text{s.t.}\quad \mathbf{M}_k \mathbf{1}_{N_{\mathbf{h}_0}} = \mathbf{h}_0,\quad \mathbf{M}_k^\top \mathbf{1}_{N_{\mathbf{h}_k}} = \mathbf{h}_k.
\end{aligned}
\end{equation}

Here, $\mathbf{M}_k \in \mathbb{R}_+^{N_{\mathbf{h}_0} \times N_{\mathbf{h}_k}}$ is the transport matrix, which is specific to each of the measured RIR $\mathbf{h}_k$ and represents how mass is optimally transported from $\mathbf{h}_k$ to $\mathbf{h}_0$. To hold the non-negative constraint of transport mass, both $\mathbf{h}_k$ and $\mathbf{h}_0$ must also be non-negative and should have an equal total mass ($||\mathbf{h}_k||_1 = ||\mathbf{h}_0||_1$).

$\mathbf{C} \in \mathbb{R}^{N_{\mathbf{h}_0} \times N_{\mathbf{h}_k}}$ represents the cost matrix and the elements of cost matrix $\mathbf{C}_{i,j}$ is chosen to be $[\mathbf{C}]_{i,j} = |\tau_i - \tau_j|^2$, where the $i$ and $j$ denote the tap indices along the sampled time axis of $\mathbf{h}_0$ and $\mathbf{h}_k$, respectively. According to the OMT principles discussed previously, if the peaks of the measured RIRs originate from the same image sources, they are expected to be closely aligned in time and have similar magnitudes which are based on their respective geometrical distances. In this paper, we aim to utilize one of the key concepts of geomatrical structure induced by delay space, called \emph{barycenter} in order to exploit the similarity between sensors located in near proximity.

Barycenter ($\mathbf{h}_0$) is used to represent the center of several mass distributions ($\mathbf{h}_k$ where $k=1,2,...K$) and in our work, the OMT barycenter minimizes the sum of the transport distances of measured RIRs to $\mathbf{h_0}$, $\operatorname{dOT}(\mathbf{h}_0, \mathbf{h}_k)$, leading us to an optimum value of $\mathbf{h_0}$ which satisfy all constraints. This can be formulated as follows,
\begin{equation}\label{eq:barycenter1}
\mathbf{h}_0 = \operatorname*{argmin}_{\mathbf{h} \ge 0} \;
\sum_{k=1}^{K} \operatorname{dOT}\bigl(\mathbf{h}, \mathbf{h}_k\bigr).
\end{equation}

Furthermore, as the equation~\eqref{eq:barycenter1} is considered convex, it is logically consistent to be utilized for the regularization of the estimation problem described in Section~\ref{sec:Proposed}. 

\section{Proposed method}\label{sec:Proposed}

In this section, we proceed to introduce the RIR estimation problem for an array of close proximity sensors. Considering the standard estimation method, least squares (LS) approach, the impulse response $\mathbf{\hat{h}}_k$ which minimizes the squared error for the signal in~\eqref{eq:measurement_matrix} can be obtained as follows,

\begin{equation}
\mathbf{\hat{h}}_k = \operatorname*{argmin}_{\mathbf{h}_k } \| \mathbf{y}_k- \mathbf{X}\mathbf{h}_k\|_2^2.
\end{equation}

If the noise  $w_k(n)$ is white Gaussian noise (WGN) with zero mean and variance $\sigma_w^2$, then minimizing the LS criterion is equivalent to obtaining the maximum likelihood estimate of the observed signal $\mathbf{y}_k$. Based on this, the LS problem is commonly defined in the literature as follows,

\begin{equation}
\operatorname*{minimize}_{\mathbf{h}_k} \| \mathbf{y}_k - \mathbf{X} \mathbf{h}_k \|_2^2.
\end{equation}


 When the length of signal is too short or the spectral excitation is too low, the signal becomes poorly conditioned as explained in Section~\ref{sigmod}. In order to obtain a unique solution for problems of this sort, we use a regularization term. Lasso ($\ell_1$) and Tikhonov ($\ell_2$)
regularization are commonly used techniques that stabilize the solution by penalizing large amplitudes. Nevertheless, while these methods provide well-defined solutions, they do not account for structural similarities in delay times across measurement points, such as the $\tau_m$ in the previously mentioned delay-magnitude pairs.

However, as discussed in Section~\ref{sec:OMT}, OMT is able to overcome these limitations by exploiting the intrinsic structure of the signal , preserving the geometrical relationship between nearby sensors and imposing a smooth delay structure.
We intend to exploit this advantage by proposing a regularization term that penalizes the OMT distance between a small group of closely positioned sensors and their shared barycenter. The barycenter serves as a \emph{virtual RIR} reference that is similar to all the measured RIRs in the microphone array. We determine this  by minimizing the collective OMT distance from each measured RIR to the barycenter, ensuring a representative and geometrically consistent reference. Furthermore, this method allows for greater ambiguity in source and receiver positions, making it a robust and more generalized regularization framework. 

Building on the discussion and utilizing information in section~\ref{sigmod}, the barycenter OMT regularization can be formulated as follows,

\begin{equation}
\operatorname*{minimize}_{\mathbf{h}_k, \mathbf{h}_0} \quad \sum_{k = 1}^K\|\mathbf{y}_k -\mathbf{X}\mathbf{h}_k  \|_2^2 + \lambda \operatorname{dOT}\bigl(\mathbf{h}_0, \mathbf{h}_k\bigr).
\label{sec:Proposedeqnonneh}
\end{equation}

Moreover, it should be noticed that, the OMT is able to naturally interpolate missing information using the calculated OMT distance. Consequently, this method can be robustly applied even when dealing with small sample sizes.

\subsection{Interpretation of RIRs in a real environment}

In our research, we aim to model the acoustic environment as realistically as possible. However, a limitation of the regularization measure discussed in Section~\ref{sec:OMT} is that, as indicated by equation~\eqref{eq:dotcal}, it is only defined for non-negative RIRs and non-negative transport plans $\mathbf{M}_k$ where $k = 1,\ldots,K$. To address this limitation and to more accurately model real environmental conditions, where both positive and negative peaks appear in RIRs due to wall absorption and distortion, we adopt the method proposed in \cite{MovingsourceRIRestDavidICASSP}. This approach decomposes the RIR into its positive and negative components, modeling their difference as follows,
\begin{equation}\label{eq:hk_decomposition}
\mathbf{h}_k = \mathbf{h}_k^+ -\mathbf{h}_k^-, \quad \mathbf{h}_k^+ \in \mathbb{R}_+^{N_\mathbf{h}}, \quad \mathbf{h}_k^- \in \mathbb{R}_+^{N_\mathbf{h}}.
\end{equation}

We can then utilize equation~\eqref{eq:hk_decomposition} to formulate a generalized OT distance between two sensors as,
\begin{equation}\label{eq:dOT_hat2}
\widehat{\operatorname{dOT}}(\mathbf{h}_1, \mathbf{h}_2) = \operatorname{dOT}\bigl(\mathbf{h}_1^+, \mathbf{h}_2^+\bigr) + \operatorname{dOT}\bigl(\mathbf{h}_1^-, \mathbf{h}_2^-\bigr).
\end{equation}

Therefore our proposed barycenter OMT regularized estimator can be reformulated based on equation~\eqref{eq:dOT_hat2} , to include both positive and negative RIRs. This formulation is given by the following equation,

\begin{equation}
\operatorname*{minimize}_{\mathbf{h}_k, \mathbf{h}_0} \quad \sum_{k = 1}^K\|\mathbf{y}_k - \mathbf{X}\mathbf{h}_k\|_2^2 + \lambda \operatorname{\widehat{dOT}}\bigl(\mathbf{h}_0, \mathbf{h}_k\bigr).
\label{sec:Proposedeq}
\end{equation}

It is worth noting that the problem in equation~\eqref{sec:Proposedeq} is a convex problem with respect to $\mathbf{h_0}$ and when negative amplitudes are concerned in the transport plan, the OMT distances may have arbitrarily scalable diagonal elements if the cost function is solely based on $| \tau_j - \tau_\ell |^2 $. In order to mitigate this issue, a small constant $\varepsilon$  can be introduced into the cost function, modifying it as $\mathbf{C}_{j,\ell} = | \tau_j - \tau_\ell |^2 + \varepsilon$. This ensures that assigning mass along the diagonal is not completely free but instead incurs a small cost.

\section{Numerical Experiments and results }

In this section, we evaluate the proposed method in \eqref{sec:Proposedeq} for estimating RIRs under different experimental setups. The simulated acoustic environment is a rectangular room with dimensions $[L_x,L_y,L_z]= 5\times 4 \times 6$ m. A human voice uttering the sustained vowel \textbackslash\textit{a}\textbackslash serves as the source signal, and a circular microphone array is used for measurements. RIRs are generated using the ISM~\cite{habets2006room}, with a sampling frequency of $f_s=7350$ Hz, a reverberation time of $\textit{RT}_{60}=0.5s$, and a maximum reflection order of 4. The length of the RIRs are all $N_h = 256$.
The source is positioned at [$2$, $3.5$, $2$] m, while the center of the microphone array is at [$2$, $1.5$, $2$] m.

%
%
The proposed method (referred to as "Barycenter OMT" in the figures) is evaluated by varying the parameters of the experimental setup. In particular, we vary the signal-to-noise ratio (SNR), the length $N_x$ of the excitation signal, the number of microphones in the microphone array, and the radius $r_0$ of the microphone array. The default values (i.e., when not varied) of the parameters are SNR = 20 dB, $N_x = 356$, $K = 5$ microphones, and $r_0 = 0.2$ meters. For the different parameter settings, we evaluate the normalized mean square error (NMSE) of the estimated RIRs, defined as
\begin{equation}
    \text{NMSE} = \frac{1}{K}\sum_{k=1}^{K}\frac{||\mathbf{h}_k-\hat{\mathbf{h}}_k||_2^2}{|\mathbf{h}_k||_2^2},
\end{equation}
where $\hat{\mathbf{h}}_k$, $k = 1,\ldots,K$, which represent the estimated RIRs. The results are averaged over 20 simulations for each setting. As reference methods, we have included the Tikhonov and Lasso estimators as well as an $\ell_2$-analog of our proposed method. The latter uses an $\ell_2$-barycenter (i.e., a standard average or mean) as a regularizer and solves
\begin{equation}
    \minwrt[\substack{\mathbf{h}_k}] \; \sum_{k=1}^{K} \Bigl\| \mathbf{y}_k - \mathbf{X}\mathbf{h}_k \Bigr\|_2^2 + \lambda \Bigl\|\mathbf{h}_k - \mathbf{h}_{\ell_{2}} \Bigr\|_2^2  +\mu\Bigl\|\mathbf{h}_k \Bigr\|_2^2,
\end{equation}
where we use the short-hand $\mathbf{h}_{\ell_{2}}=\frac{1}{K}\sum_{k=1}^{K}\mathbf{h}_k$. It may be noted that we have included a small Tikhonov-type term for this reference as it may be verified that the problem is otherwise ill-conditioned when $\mathbf{X}$ is ill-conditioned. We refer to this method as "$\ell_2$" in the results.
Also, as a reference, we include an array version of the method from~\cite{MovingsourceRIRestDavidICASSP}, which penalizes the OMT distance between the RIRs of adjacent microphones clockwise around the microphone array.
%
%
This reference is indicated "Adjacent OMT" in the graphs, which solves
%
%
\begin{equation}\label{eq:adjacent OMT}
\minwrt[\mathbf{h}_k] \quad \sum_{k = 1}^K\|\mathbf{y}_k - \mathbf{X}\mathbf{h}_k \|_2^2 + \lambda 
\sum_{\ell=2}^{K} \operatorname{\widehat{dOT}}\bigl(\mathbf{h}_\ell, \mathbf{h}_{\ell-1}\bigr).
\end{equation}
For all methods, the values of their respective regularization parameters are set by cross-validation.

\begin{figure}[t]
\centering
\subfloat{\includegraphics[width=0.45\textwidth]{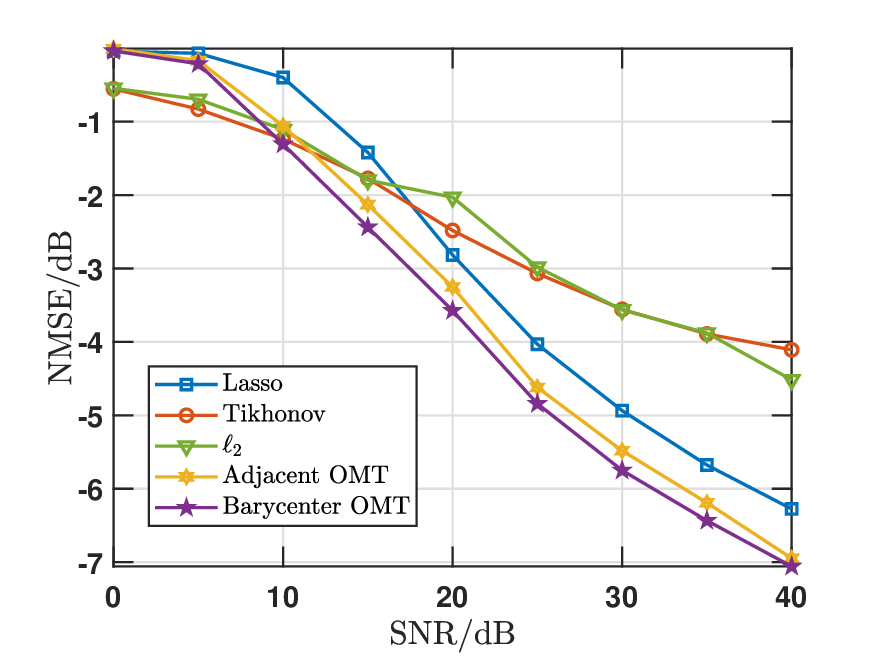}}
\caption{Simulated estimation error $\text{NMSE}$ for different regularization methods as a function of the SNR.}
\label{Numerical:SNR}
\end{figure}

\begin{figure}[t]
\centering
\subfloat{\includegraphics[width=0.45\textwidth]{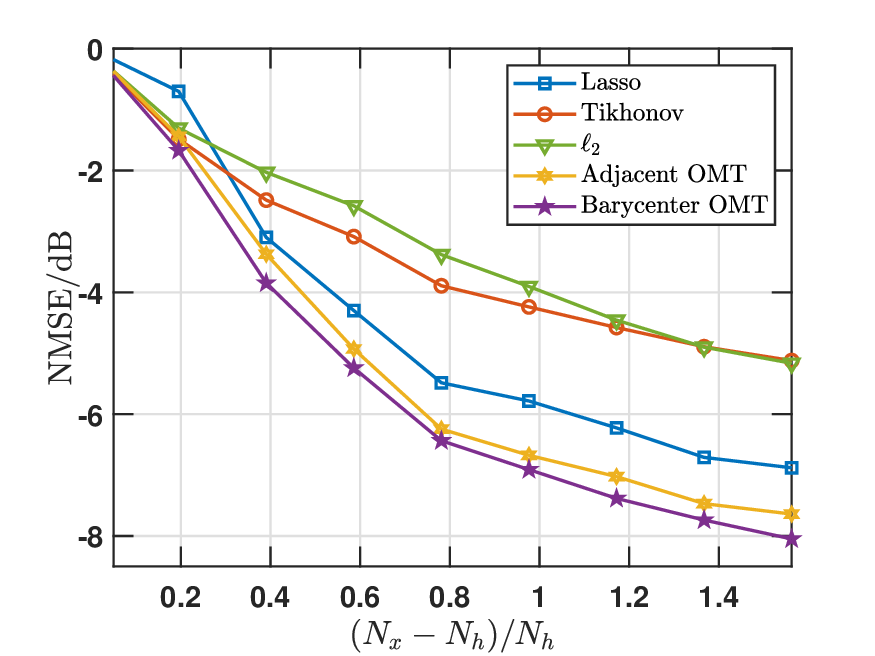}}
\caption{Simulated estimation error $\text{NMSE}$ for different regularization methods as a function of the $(N_x-N_h)/N_h$.}
\label{Numerical:Nx}
\end{figure}
\begin{figure}[tp]
\centering
\subfloat{\includegraphics[width=0.45\textwidth]{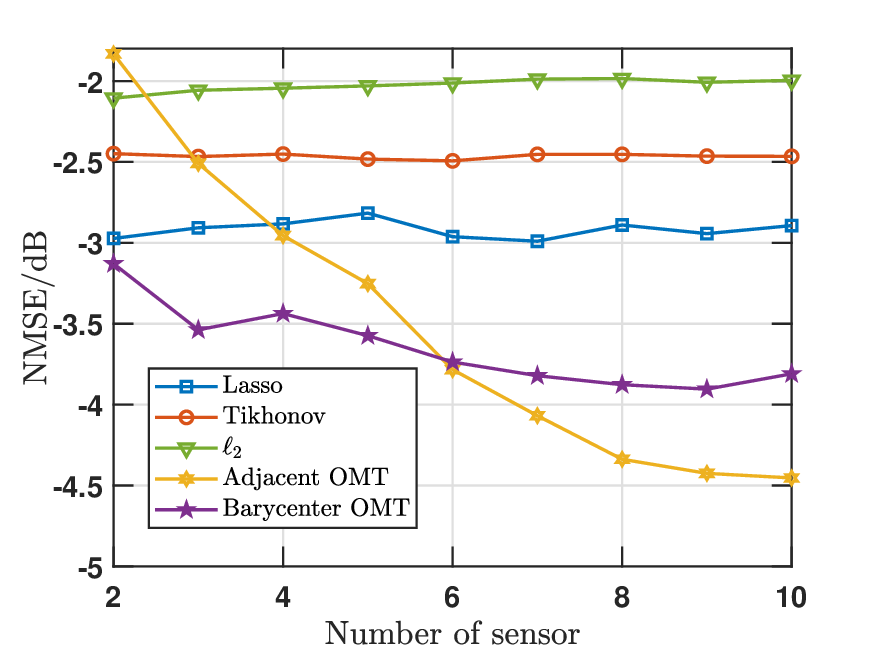}}
\caption{Simulated averaged estimation error $\text{NMSE}$ for different regularization methods as a function of the number of sensors.}
\label{Numerical:nbsensor}
\end{figure}
\begin{figure}[tp]
\centering
\subfloat{\includegraphics[width=0.45\textwidth]{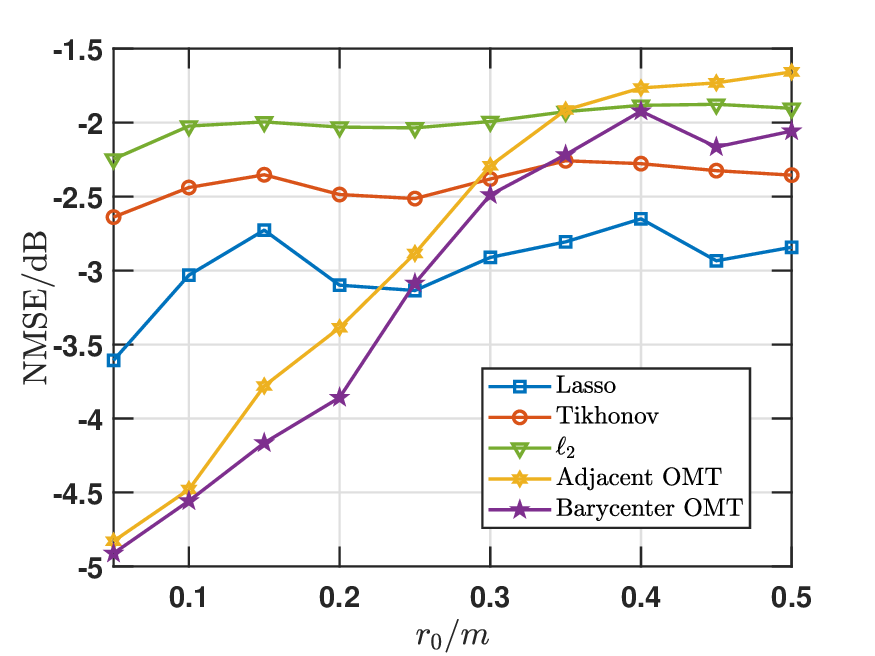}}
\caption{Simulated averaged estimation error $\text{NMSE}$ for different regularization methods as a function of the radius of microphone array $r_0$.}
\label{Numerical:r0}
\end{figure}

Figures~\ref{Numerical:SNR} and \ref{Numerical:Nx} present the NMSE (in dB) for varying SNR and excitation signal length $N_x$, respectively. Note in Figure~\ref{Numerical:Nx}, the x-axis is the ratio $(N_x-N_h)/N_h$. As observed, the proposed method performs better than the comparison methods, although Lasso and the Adjacent OMT offers similar performance. However, when varying the number of microphones in the array and the array's radius, we see a remarkable difference between the OMT-based methods and the other three references. These results can be seen in Figures~\ref{Numerical:nbsensor} and \ref{Numerical:r0}, respectively. The results from figure~\ref{Numerical:nbsensor} indicate that the performance of the non-OMT references are independent of the number of microphones. This is expected for the Lasso and Tikhonov estimators, as they do not account for information sharing between microphones when estimating the RIRs. The $\ell_2$ also doesn't benefit from an increasing number of microphones, as it is built on Euclidean geometry that is not designed to offer a good description of variation of RIRs  across the array. On the other hand, the OMT-based methods, i.e., both the proposed estimator and reference, are able to exploit the information provided by an increasing number of sensors. This observation is even more pronounced in figure~\ref{Numerical:r0}. Here, as the radius of the array shrinks, i.e., as the microphones become closer to each other, the error for the OMT-based methods decrease. This is due to the fact that as the microphones become more closely spaced, their corresponding RIRs become more similar in the OMT sense. Conversely, if the array radius becomes too large, the performance of the OMT-based methods drop.
%



\section{Conclusion}


This paper introduces a barycenter-based OMT regularization method for estimating RIRs when the excitation signal is poorly informed. Unlike conventional approaches that independently regularize each RIR or only consider pairwise adjacent RIRs, the proposed technique couples all sensor RIRs through a common virtual barycenter, effectively capturing similar delay structures in spatially proximate receivers, enhancing robustness in small measurement areas with limited sensors. Numerical experiments confirm that the method outperforms baseline estimators and surpasses adjacent sensor OMT in scenarios with low sensor counts. Future work will focus on adaptive barycenter modeling to interpolate RIRs, incorporating physically informed cost functions.


\bibliographystyle{ieeetr}
\bibliography{ref.bib}

\end{document}